\documentclass[useAMS,usenatbib]{mn2e}
\usepackage{epsfig}
\usepackage{amsmath, amssymb,bm}
\usepackage[varg]{txfonts}
\usepackage{color}
\usepackage{aas_macros}
\usepackage[export]{adjustbox}
\usepackage{pdfpages}

\title[Frequency of giant outbursts in Be/X-ray binaries]{The frequency of Kozai--Lidov disc oscillation  driven giant outbursts   in Be/X-ray binaries}

\author[R. G. Martin \& A. Franchini]{Rebecca G. Martin\thanks{E-mail:
    rebecca.martin@unlv.edu} and Alessia Franchini\\Department
  of Physics and Astronomy, University of Nevada, Las Vegas, 4505
  South Maryland Parkway, Las Vegas, NV 89154, USA \\ }

\date{}

\pagerange{\pageref{firstpage}--\pageref{lastpage}} 
\pubyear{2019}

\begin{document}
\maketitle
\label{firstpage}
\begin{abstract} 
Giant outbursts of Be/X-ray binaries may occur when a Be--star disc undergoes strong eccentricity growth due to the Kozai--Lidov (KL) mechanism.  The KL effect acts on a disc that is highly inclined to the binary orbital plane provided that the disc aspect ratio is sufficiently small. The eccentric disc overflows its Roche lobe and material flows from the Be star disc over to the companion neutron star causing X-ray activity.   With N--body simulations and steady state decretion disc models we explore system parameters for which a  disc in  the Be/X-ray binary 4U~0115+634 is KL unstable and the resulting timescale for the oscillations. We find good agreement between predictions of the model and the observed giant outburst timescale provided that the disc is not completely destroyed by the outburst. This allows the outer disc to be  replenished between outbursts and a sufficiently short KL oscillation timescale. An initially eccentric disc has a shorter KL oscillation timescale compared to an initially circular orbit disc.     We suggest that the chaotic nature of the outbursts is caused by the sensitivity of the mechanism to the distribution of material within the disc. The outbursts continue provided that the Be star supplies material that is sufficiently misaligned to the binary orbital plane.  We generalise our results to Be/X-ray binaries with varying orbital period and find that if the Be star disc is flared, it is more likely to be unstable to KL oscillations in  a smaller orbital period binary, in agreement with observations.
\end{abstract} 
  
\begin{keywords} 
accretion, accretion disks -- binaries: general -- stars: emission-line, Be -- pulsars: individual (4U~0115+634) -- X- rays: binaries
\end{keywords} 
 
\section{Introduction}   
 
Be stars are rapidly rotating early type stars with a viscous thin,
Keplerian disc
\citep{Lee1991,Porter1996,Hanuschik1996,Hummel1998,Quirrenbach1997,Porter2003,Rivinius2013,Okazaki2016}. Material
is added to the inner parts of the disc from the Be star
\citep{Cassinelli2002} and thus the disc is a decretion disc, rather
than an accretion disc \citep{Pringle1991}. The stellar material spirals outwards
through the disc due to the action of viscosity that is thought to be
driven by the magnetorotational instability \citep{BH1991}.

Be/X-ray binaries typically consist of a Be star and a neutron star in an
eccentric orbit binary. The orbit became eccentric, and most likely
inclined with respect to the spin of the Be star, when the neutron
star formed in a supernova explosion. A slight asymmetry to the
explosion caused a kick on the newly formed neutron star
\citep{Brandt1995,Martinetal2009b}. The tidal torque exerted by the neutron star companion
truncates the Be star disc  \citep{Okazaki2002,Hayasaki2004,Reig2007b}.
This is observationally confirmed since the disc is more dense when
the Be star has a binary companion \citep{Zamanov2001,Cyr2017}.

Some Be/X-ray binaries are transient systems, meaning that they alternate between long quiescent periods and short outbursts, while others are persistent X-ray sources \citep{Okazaki2002}.
There are two types of X-ray outbursts that occur through accretion
on to the neutron star \citep[e.g.][]{Kuehnel2015}. Type~I X-ray
outbursts in an eccentric orbit binary occur each orbital period when the neutron star is at
periastron and able to capture material from the Be star disc
\citep{Okazaki2001,Negueruela2001,Okazaki2013}. In a low eccentricity binary, type~I outbursts may be driven by a disc that becomes eccentric through mean motion resonances with the binary \citep{Franchini2019c}. Type~II (or giant) outbursts  are much brighter, last for longer and occur
less frequently \citep{Stella1986,
  Negueruela1998,Kretschmar2013,Cheng2014,Monageng2017}.
  
  In this work we focus on the giant outbursts and in particular the Be/X-ray binary 4U~0115+634. This object is one of the first discovered \citep{Giacconi1972,Whitlock1989} and best-studied Be X-ray binaries \citep[e.g.][]{Campana1996,Negueruela1997}.  This a classical system with well constrained orbital parameters on which many models for the behaviour of Be X-ray binaries are based \citep[e.g.][]{Negueruela2001b,Okazaki2001}.   
 
 Giant outbursts in this system have been associated with the presence of a tilted disc around the Be star
\citep{Reigetal2007,
  Moritanietal2011,Martinetal2011,Moritani2013, Kato2014}.   The observed timescale
between giant outbursts is around $3\,\rm yr$
\citep{Whitlock1989,Negueruela2001,Reigetal2007,Reig2018}. However,
sometimes these outbursts are separated by only 1-1.5$\,\rm yr$, and in this
case the second outburst is of lower luminosity than the first \citep[e.g.][]{Rauco2017,Rauco2019}. After
major outbursts a decrease in the size of the Be star disc is observed
\citep{Reigetal2007,Reig2016}.

A sufficiently misaligned disc around a Be star can become highly eccentric
\citep{Martinetal2014}. This is a result of Kozai--Lidov oscillations
that exchange inclination and eccentricity of a misaligned 
orbit around one component of a binary
\citep{Kozai1962,Lidov1962}. In a fluid disc, if the radial sound
crossing timescale is shorter than the period of the KL oscillations
then the disc can undergo global KL oscillations
\citep{Martinetal2014b,Fu2015,Fu2015b,Fu2017}. The eccentricity growth causes
the disc to overflow its Roche lobe and transfer material to the
companion neutron star \citep{Franchini2019}. This may be the
mechanism behind type II X-ray outbursts observed in Be/X-ray
binaries.

In this work we consider the frequency of type II outbursts as a
result of KL oscillations of the Be star disc.  In
Section~\ref{testparticles} we first consider the timescale for KL
oscillations of a test particle orbit around the Be star in the binary
4U~0115+634.  In Section~\ref{discmodel} we explore the KL
oscillation timescale for a steady state Be star decretion disc  around the Be star in the binary
4U~0115+634. We examine parameters for which the disc is KL unstable and find that our model agrees well with the numerical simulation presented in \cite{Martinetal2014}. The advantage of our analytic approach is that we can examine a large parameter space. In Section~\ref{porb} we  generalise our results to a wider range of parameters and show that giant outbursts driven by KL oscillations are more likely to occur for smaller orbital period binaries. In
Section~\ref{discussion} we discuss the implications of our results
and we draw our conclusions in Section~\ref{concs}.

\begin{figure} 
\begin{centering} 
\includegraphics[width=8.5cm]{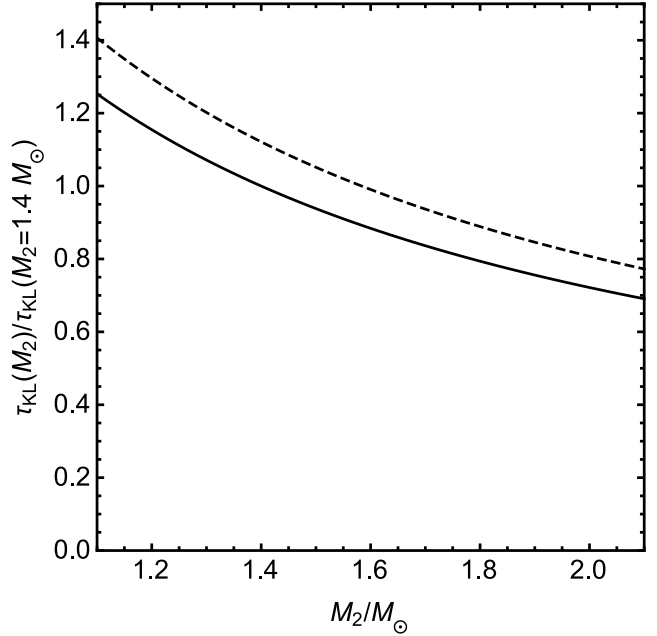} 
\end{centering} 
\vspace{-0.8cm}
\caption{ The relative change in the KL oscillation timescale of a particle with varying neutron star companion mass, $M_2$. The solid line has a Be star of mass $M_1=18\,\rm M_\odot$ while the dashed line has  $M_1=19.5\,\rm M_\odot$.   }
\label{fig:mass} 
\end{figure} 

\begin{figure*} 
\includegraphics[width=8.5cm,valign=t]{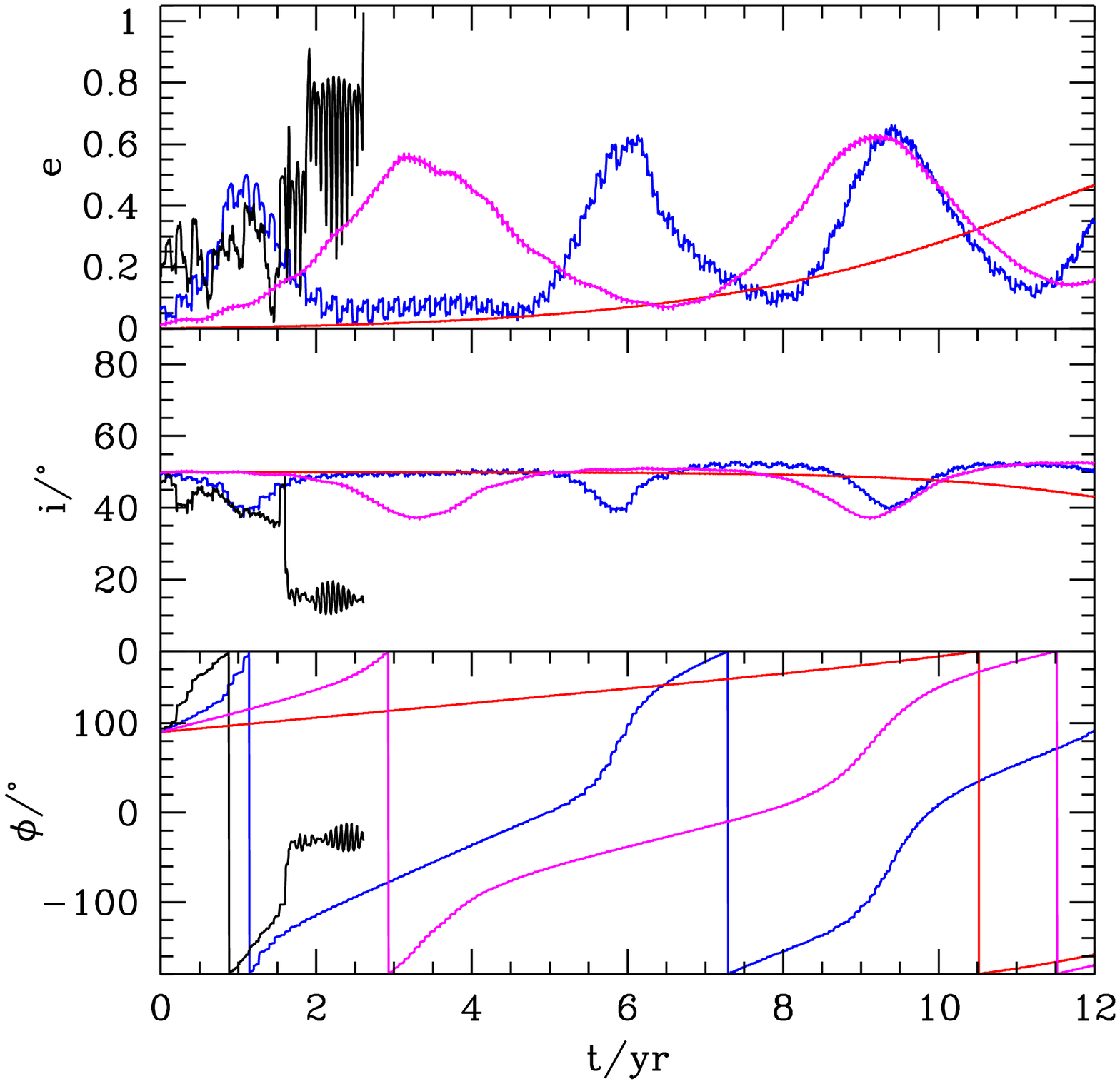} 
\includegraphics[width=8.5cm,valign=t]{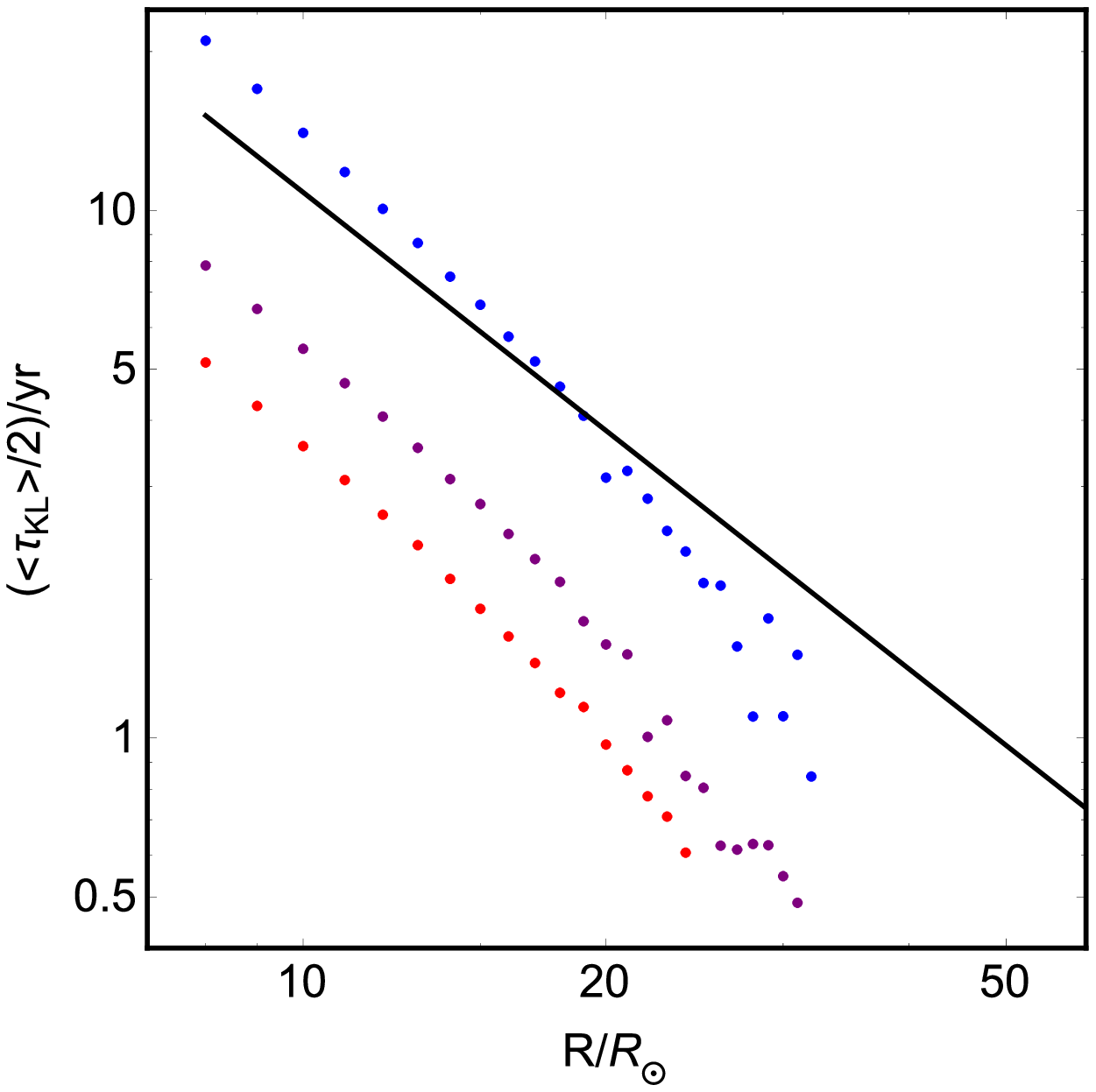} 
\caption{Left: Eccentricity, inclination and phase angle for test
  particle simulations around the Be star. The orbital plane is
  initially misaligned by $50^\circ$ with phase angle $\phi=90^\circ$.
  The orbits are at orbital separation $10\,\rm R_\odot$ (red),
  $20\,\rm R_\odot$ (magenta), $30\,\rm R_\odot$ (blue) and $40\,\rm
  R_\odot$ (black).  Right: The time of the first KL oscillation
  eccentricity peak for the analytic prediction
  (equation~\ref{time}). The points show the time of the first peak as a function of test particle semi--major axis for simulations
  that are initially misaligned by $50^\circ$ with $\phi=90^\circ$. The blue points are initially circular test particle orbits, the purple points have initial particle eccentricity $e_{\rm p}=0.2$ and the red points have $e_{\rm p}=0.4$.}
\label{fig:test2} 
\end{figure*}

\section{KL oscillation timescale for a particle orbit}
\label{testparticles}

We consider a binary with components of mass $M_1$ and $M_2$, orbital
period $P_{\rm orb}=2\pi/\sqrt{G(M_1+M_2)/a^3}$,  (where $G$ is the gravitational constant), semi--major axis,
$a$, and orbital eccentricity $e_{\rm b}$. The KL effect causes
oscillations of the inclination, $i$, and eccentricity, $e$, of a
misaligned test particle around one component of a binary system
\citep{Kozai1962,Lidov1962}. For an initially circular test particle
orbit around one component of a circular orbit binary, oscillations
occur when the initial inclination satisfies $39^\circ \lesssim i
\lesssim 141^\circ$. During the oscillations, the component of the
angular momentum of the particle that is perpendicular to the binary
orbital plane is conserved so that
\begin{equation}
\sqrt{1-e^2}\cos i\approx \rm const.
\end{equation}
The KL oscillation timescale is  approximately
\begin{equation}
\frac{\tau_{\rm KL}}{P_{\rm orb}}\approx\frac{M_1+M_2}{M_2}\frac{P_{\rm orb}}{P_{\rm p}}(1-e_{\rm b}^2)^\frac{3}{2},
\label{time}
\end{equation}
where $P_{\rm p}=2\pi/\sqrt{GM_1/R^3}$ is the orbital period of
the particle that orbits around the Be star with semi-major axis $R$.  We consider here first how properties of the binary orbit might change the KL oscillation timescale and then how properties of the particle orbit change the timescale.

\subsection{Binary parameters}
\label{binary}

We consider a Be star binary model based on the parameters of 4U~0115+634. For our standard model we follow \cite{Negueruela2001} and take the Be star to have a mass $M_1=18\,\rm M_\odot$  and radius $R_1=8\,\rm R_\odot$  \citep{Vacca1996}. The neutron star companion has a mass
$M_2=1.4\,\rm M_\odot$ and is at an orbital semi--major axis of
$a=95\,\rm R_\odot$ with orbital eccentricity $e=0.34$  \citep{Rappaport1978}. The orbital
period is $P_{\rm orb}=24.3\,\rm day$.  The errors in the orbital properties of 4U~0115+634 are very small. The orbital eccentricity is $0.3402\pm0.0004$ and the orbital period $P_{\rm orb}=24.309\pm 0.021\,\rm day$ \citep{Rappaport1978}. The masses we take for each component in our model are on the low end of the possible ranges. The mass of the Be star is determined through its spectral type (B0.2Ve) and could be up to about $19.5\,M_\odot$ \citep{Vacca1996}. The mass of a neutron star is in the range $1.1-2.1\,\rm M_\odot$ \citep[e.g.][]{Ozel2012}.

Fig.~\ref{fig:mass} shows the change in the KL oscillation timescale (calculated with equation~(\ref{time}), for fixed binary orbital period and for a particle of fixed semi--major axis) as a function of the neutron star mass.  The two lines show different Be star masses, $M_1=18\,\rm M_\odot$ (solid line) and $M_1=19.5\,\rm M_\odot$ (dashed line). The possible error in the Be star mass does not significantly affect the KL oscillation timescale. However, the error in the unknown mass of the neutron star could lead to a decrease in the KL oscillation timescale of up to a maximum of about $30\%$. We discuss this further in Section~\ref{KL}.

\subsection{Test particle orbital parameters}
\label{test}

We consider here test particle
orbits around the Be star with varying orbital properties. The left hand panel of Fig.~\ref{fig:test2} shows test particle orbits
around the Be star initially inclined by $50^\circ$ and starting at
different orbital separations. The binary orbits in the $x-y$ plane and
begins at periastron along the $x$ axis. Since the particle is
massless the binary orbit remains fixed. The direction of the angular
momentum of the particle is given by the unit vector
$\bm{l}=(l_x,l_y,l_z)$ and its phase angle is defined as
\begin{equation}
\phi=\tan^{-1}\left(\frac{l_y}{l_x}\right).
\end{equation}
The initial phase angle is $\phi=90^\circ$.  Particles closer to the
Be star have a longer KL oscillation timescale, as predicted by
equation~(\ref{time}). The farther out a particle is, the more chaotic
its evolution. A particle that begins at a separation of $40\,\rm
R_\odot$ is ejected from the system. 
The right hand panel of
Fig.~\ref{fig:test2} shows the analytic prediction for the time of the first
peak eccentricity given in equation~(\ref{time}). The blue points show the
results of test particle simulations with an initially circular orbit
inclined by $i=50^\circ$ with $\phi=90^\circ$. We note that the oscillation timescale is not uniform 
because of the eccentricity of the binary \citep[e.g.][]{Naoz2013}. Also, the orbits we consider are not in the hierarchical triple body limit in which the separation of the particle from
the Be star is much smaller than the separation of the binary. 

An initially eccentric particle orbit has a shorter KL oscillation timescale \citep{Franchini2019}. In the right hand panel of Fig.~\ref{fig:test2}, the purple points show particles with initial eccentricity $e_{\rm p}=0.2$ and the red points with $e_{\rm p}=0.4$. The initial inclination is again $50^\circ$, the  phase angle is $\phi=90^\circ$ and the argument of periapsis is $180^\circ$.  For small separation, the KL oscillation timescale is a factor of 4.1 shorter for an initial eccentricity of 0.4 and a factor of 2.7 shorter for initial eccentricity of 0.2. 

\begin{figure} 
\begin{centering} 
\includegraphics[width=8.5cm]{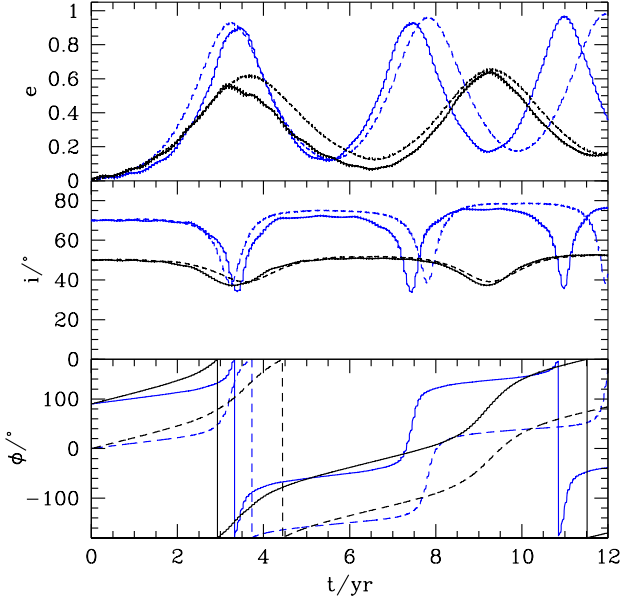} 
\end{centering} 
\caption{ The
  eccentricity (upper), inclination (middle) and phase angle
  (lower) of} test particle orbits that are initially circular and inclined
  by $50^\circ$ (black lines) and $70^\circ$ (blue lines) to the
  binary orbital plane at semi--major axis $20\,R_\odot$.  The solid lines show an orbit with $\phi=90^\circ$ initially
  and the dashed lines with $\phi=0^\circ$ initially.  
\label{fig:test} 
\end{figure}

Fig.~\ref{fig:test} shows test particle simulations that begin at
different nodal phases, $\phi=0^\circ$ and $\phi=90^\circ$ for two
different initial inclinations. These show that the time to the first
peak is relatively insensitive to the initial inclination or the nodal
phase angle. We also found that the KL oscillation timescale is rather insensitive to the initial argument of periapsis.  The parameters that the timescale is most sensitive to are
the semi--major axis of the particle and the initial eccentricity of the particle orbit.

For a circular orbit disc composed of test particles, each particle undergoes KL
oscillations and nodal precession on a timescale that most strongly depends on its
separation from the Be star. The disc of particles forms a thick structure or swarm on a short timescale. However, the pressure in a hydrodynamical gaseous
disc connects different orbital radii of the disc together allowing it to undergo
global KL oscillations. The decretion disc can extend to larger radii than those of stable test particle orbits.
The timescale of the oscillations depends upon the distribution of
material within the disc. In the next Section we explore the KL
oscillation timescale for a gas disc.

\section{Decretion disc model}
\label{discmodel}

In this Section we first consider the Be star decretion disc temperature structure and the surface density profile.  We focus on the orbital parameters of the Be/X-ray binary 4U~0115+634. We estimate the global KL oscillation timescale for a Be star decretion  disc as a means to calculate the frequency of type II X-ray outbursts.  As the disc undergoes a KL oscillation, it becomes very eccentric. However, the inner parts of the disc continue to accrete circular orbit material from the star. The interaction of the circular orbit material
being added to an eccentric disc is beyond the scope of this work but
will be investigated in the future. Here, we consider a steady state
decretion disc model for the Be star disc. This represents the quasi--steady state disc that is truncated by the tidal torque from the companion neutron star.

\subsection{Disc temperature structure}
\label{tvisc}

The material in the disc orbits the Be star at Keplerian angular
frequency $\Omega=\sqrt{G M_1/R^3}$. The disc extends from the stellar
radius, $R_{\rm in}=8\,\rm R_\odot$, up to $R_{\rm t}$. 
The outer radius of the disc, $R_{\rm t}$, is determined by  where the disc is tidally truncated by the companion
neutron star.   The outer edge of the disc for 4U~0115+634 in the smoothed particle hydrodynamical simulations of \cite{Martinetal2014} is $R_{\rm t}\approx 50\,\rm R_\odot$. 
The viscosity of the disc is given by
\begin{equation}
\nu=\alpha \left(\frac{H}{R}\right)^2 R^2 \Omega,
\end{equation}
where $H$ is the disc scale height and $\alpha$ is the \cite{SS1973}
viscosity parameter. The value of $\alpha$ in hot Be star discs is
likely around $0.3$
\citep{Jones2008,Carciofi2012,Ghoreyshi2017,Rimulo2018}. This is
typical for a fully ionised accretion disc in which the viscosity is
driven by the magneto--rotational instability \citep{Martin2019}.
We assume that the aspect ratio is a power law in radius
\begin{equation}
\frac{H}{R}=\left(\frac{H}{R}\right)_{0} \left(\frac{R}{R_{0}}\right)^s,
\label{eq:aspectratio}
\end{equation}
where $s$ is a constant and $(H/R)_{0}$ is the disc aspect ratio at $R=R_0$. We choose to scale the disc aspect ratio at an orbital radius of $R_{0}=50\,\rm R_\odot$.  
Fig.~\ref{sigma} shows the disc aspect ratio  for  three values for the power index,  $s=0$  (a constant disc aspect ratio), $s=0.25$  and $s=0.5$ (an isothermal disc).

Observationally, the disc aspect ratios of Be stars discs are not well constrained. \cite{Wood1997} found the aspect ratio at the inner edge of $\zeta$ Tauri to be 0.04 while other estimates suggest the disc aspect ratio may be much larger farther out in the disc \citep[e.g][]{Porter1996,Quirrenbach1997}. \cite{Hanuschik1996} found that the disc flares at large radii.
\cite{Negueruela2001} modelled the inner disc edge to be $0.026$ for 4U~0115+634. Due to these uncertainties we have chosen to parameterise our models by the disc aspect ratio at the outer edge since this is more important for the dynamical evolution of the disc.  Given the uncertainty in the disc aspect ratio, we consider two different values of $(H/R)_{0}=0.1$ and $(H/R)_{0}=0.06$. 

\subsubsection{Criteria for the disc to be KL unstable}

\begin{figure} 
\begin{centering} 
\includegraphics[width=8.5cm]{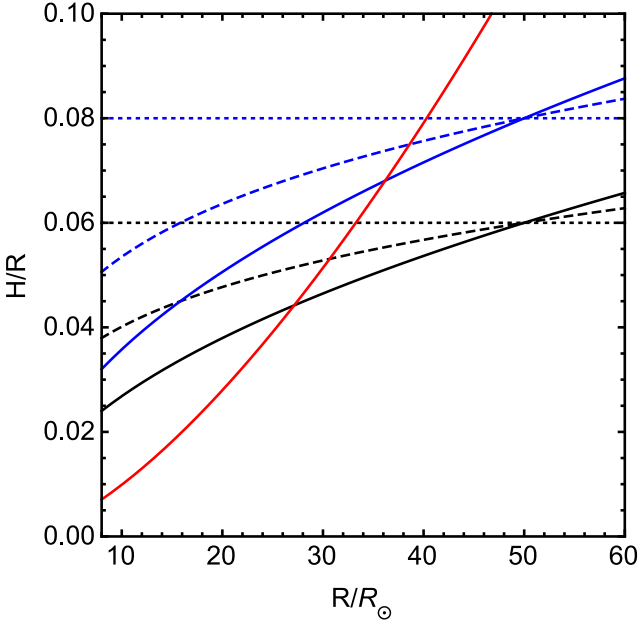} 
\end{centering} 
\vspace{-0.5cm}
\caption{The disc aspect ratio as a function of radius for $s=0.5$ (solid lines), $s=0.25$ (dashed lines) and $s=0$ (dotted lines). The blue (upper) lines have $(H/R)_{0}=0.08$ and the black (lower) lines have $(H/R)_{0}=0.06$ with $R_{0}=50\,\rm R_\odot$. The solid red line shows the critical value of the disc aspect ratio in the outer parts of the disc below which the disc is KL unstable,  $(H/R)_{\rm crit}$ (equation~\ref{hcrit}). }
\label{fig:sigma} 
\end{figure}

The KL oscillation of a disc may be suppressed if the value for $H/R$ at the disc outer edge becomes too large \citep{Lubow2017,Zanazzi2017}. For a small mass perturber, the critical value at the disc outer edge is given approximately by
\begin{equation}
\left(\frac{H}{R}\right)_{\rm crit}=\frac{\sqrt{M_2 M}}{M_1}\sqrt{\frac{R^3}{a^3}}
\label{hcrit}
\end{equation}
\citep{Lubow2017}, where $M=M_1+M_2$ is the total mass of the binary.
For a radius of $R=50\,\rm R_\odot$, we find $(H/R)_{\rm crit}=0.11$. For the parameters of 4U~0115+634, we plot the critical aspect ratio as a function of  radius in the solid red line in Fig.~\ref{fig:sigma}. Since the disc models we consider all have smaller disc aspect ratio in the outer parts of the disc, they are unstable to KL disc oscillations. 

The critical disc aspect ratio estimate for the KL mechanism to operate is in agreement with simulations of isothermal discs published in \citet{Martinetal2014}. We found that the KL mechanism did not operate for $H/R(R_{\rm in})=0.04$, but operated for smaller values considered. This corresponds to disc aspect ratio at the outer edge of about  $H/R(R=50\,\rm R_\odot)=0.1$. Thus the numerical simulations are in good agreement with the analytic estimate in equation~(\ref{hcrit}). The two isothermal disc models shown in Fig.~\ref{fig:sigma} (solid blue and black lines) both have $H/R(R_{\rm in})<0.04$.

\subsubsection{Disc viscous timescale}

\begin{figure} 
\begin{centering} 
\includegraphics[width=8.5cm]{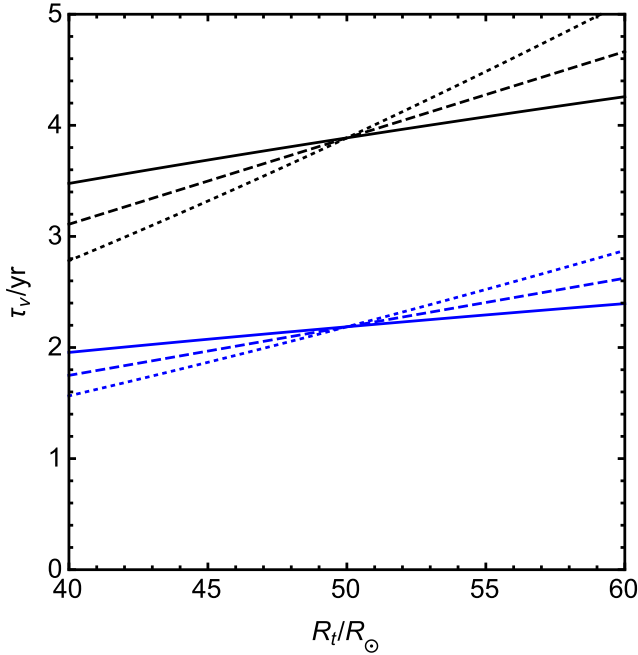} 
\end{centering} 
\vspace{-0.5cm}
\caption{Viscous timescale of a Be star decretion disc according to
  equation~(\ref{eq:tnu})  for $s=0.5$ (solid lines), $s=0.25$
  (dashed lines) and $s=0$ (dotted lines). The blue (lower) lines have $(H/R)_{0}=0.08$ and the black (lower) lines have $(H/R)_{0}=0.06$ with $R_{0}=50\,\rm R_\odot$. }
\label{fig:tnu} 
\end{figure}

The viscous timescale at the outer edge of the disc is defined as $\tau_\nu = R_{\rm t}^2/\nu$, where $R_{\rm t}$ is the outer truncation radius for the disc. For typical parameters this is given by
\begin{equation}
\tau_\nu = 2.2\, \left(\frac{\alpha}{0.3}\right)^{-1} \left(\frac{(H/R)_{\rm t}}{0.08}\right)^{-2}\left(\frac{M_1}{18\,\rm M_\odot}\right)^{-\frac{1}{2}}\left(\frac{R_{\rm t}}{50\,\rm R_\odot}\right)^{\frac{3}{2}-2s} 
\,\rm yr.
\label{eq:tnu}
\end{equation}
 Fig.~\ref{fig:tnu} shows the viscous timescale as a function of radius for varying $s$.  Since the outer parts of the disc are
 depleted during the outburst, material needs to spread
outwards on a timescale shorter than the type~II outburst timescale. 
This requires the disc aspect ratio at the outer disc edge to be $H/R\gtrsim 0.06$.

We have constrained the disc aspect ratio at the outer edge of the disc to be in the range $0.06 \lesssim H/R \lesssim 0.11$. Even the shortest possible viscous timescale (for small outer truncation radius and $(H/R)_{0}=0.1$) is only about a factor of $1.5-3$ times smaller than the outburst frequency. Thus, we suggest that the disc cannot be completely depleted during an outburst in order to have such frequent outbursts.

\subsection{Surface density of the disc}

For a circular orbit binary, test particle orbits cross at a
distance of about $0.48\,a =45.6\,\rm R_\odot$
\citep{Paczynski1977} and this corresponds to the maximum radius at which a cold disc can exist. However a Be star disc may extend to larger radii than this as
it is stabilised by pressure effects within it. The
eccentricity of the binary orbit works in the opposite direction and
leads to a smaller disc \citep{Artymowicz1994}. However, an inclined
disc feels a weaker binary torque than a coplanar disc and thus may be
larger \citep{Lubowetal2015}. Hydrodynamical simulations of such a
disc find that the tidal truncation radius varies in time, but for a disc that is initially misaligned by $70^\circ$, it is typically around $50\,\rm R_\odot$ \citep[][]{Martinetal2014}. The tidal torque increases very strongly with radius \citep[e.g.][]{Papaloizou1977,MartinandLubow2011} and the surface density profile is sharply truncated at the outer disc edge.

In this work, we do not
explicitly include the tidal torque from the neutron star that truncates the outer edge of the disc but the surface density profile is truncated at radius $R_{\rm t}$. 
For the Be star decretion
disc the steady state surface density is
\begin{equation}
\Sigma=\Sigma_0 \left(\frac{R}{R_{\rm in}}\right)^{-n+s+1}
\label{sigma}
\end{equation}
for $R_{\rm in}<R<R_{\rm t}$, where $\Sigma_0$ is a constant that is determined by the total mass of
the disc and $n$ is a constant. We chose this power law so that the density scales as $\rho \propto \Sigma/ H \propto R^{-n}$. The value for $\Sigma_0$ changes in time depending on the
total mass of the disc \citep[e.g.][]{Bjorkman2005,Carciofi2006}. 
The amount of mass contained in the disc depends upon the accretion rate of material from
the Be star to the disc and the mass loss rate from the
disc. 
Since the disc KL oscillation
timescale does not depend upon the total mass of the disc, but rather
the distribution of mass, we do not make any assumption on the absolute
value of the surface density of the disc. 

The surface density of a steady state decretion disc satisfies $\nu \Sigma \propto R^{-1/2}$ \citep[e.g.][]{Pringle1981,Martinetal2011}. For the disc to be steady
we require the power law indices to be related by
\begin{equation}
n=3s+2
\label{pq}
\end{equation}
The more strongly flared the disc vertical structure, the sharper the drop off with radius in the surface density profile of the steady state disc.

Observationally, in the inner parts of Be star discs, the density at the midplane scales as $\rho \propto R^{-n}$ where $n$ is in the range $2-3.5$ for an isothermal disc \citep[e.g.][]{Cote1987,Porter1999}.  The sound speed is $c_{\rm s}=H \Omega$ and thus an isothermal disc has $H/R \propto R^{1/2}$, or $s=0.5$. With equation~(\ref{pq}), the steady state isothermal disc has $n=3.5$,  in agreement with the observations.

\subsection{KL oscillation timescale}
 \label{KL}

 If the KL disc mechanism drives giant outbursts, then the KL oscillation timescale for the disc corresponds to
the  timescale between subsequent type II
outbursts. For an initially circular test particle orbit, the eccentricity may go to zero between oscillations. However, because of dissipation within the disc, the eccentricity oscillation is damped and does not go all the way back to zero. Thus, if there is significant remaining material in the disc after an outburst, it may have high
eccentricity without filling the Roche lobe and transferring to the neutron star. 
The KL oscillation timescale is shorter for test particles that start on an already eccentric orbit (see Figure~\ref{fig:test2}). Thus, the KL oscillation timescale estimates in this Section are
upper limits to the  predicted timescale between type II outbursts.

The timescale for global KL oscillations of a disc is approximated by
\begin{equation}
\left<\tau_{\rm KL}\right>=\frac{\int_{R_{\rm in}}^{R_{\rm t}}\Sigma R^3 \sqrt{\frac{GM_1}{R^3}}\,dR  }{\int_{R_{\rm in}}^{R_{\rm t}}\tau_{\rm KL}^{-1}\Sigma R^3 \sqrt{\frac{GM_1}{R^3}}\,dR}
\end{equation}
\citep{Martinetal2014b}.  The total mass of the disc does not affect
the timescale for KL oscillations.  The KL oscillation timescale is
sensitive to the size of the disc and the distribution of
material within it. We consider the steady state Be star disc surface
density solution given in equations~(\ref{sigma}) and~(\ref{pq}).

\begin{figure} 
\begin{centering} 
\includegraphics[width=8.5cm]{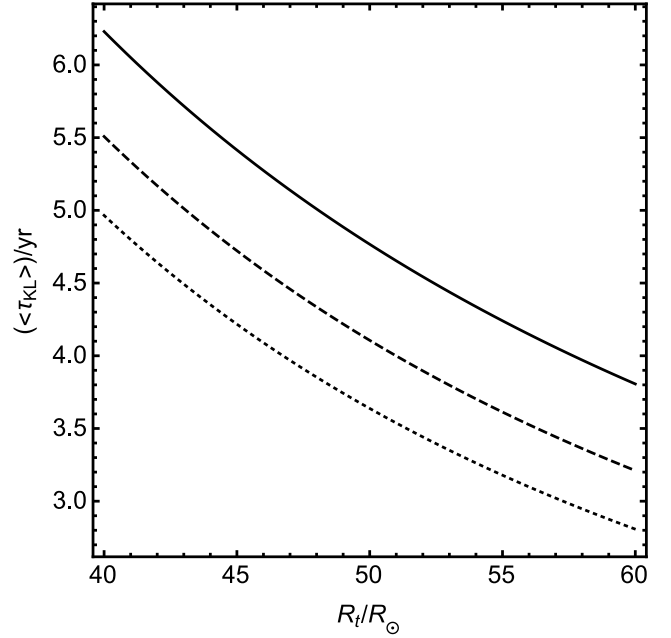} 
\end{centering} 
\caption{The KL oscillation timescale for an initially circular steady state decretion disc with
  surface density distributed as equation~(\ref{sigma}) between the
  stellar radius $R_{\rm in}=8\,\rm R_\odot$ and $R_{\rm t}$ for
  $s=0.5$ (solid line), $s=0.25$ (dashed line) and $s=0$ (dotted line).}
\label{fig:kl} 
\end{figure}

Fig.~\ref{fig:kl} shows the KL oscillation timescale for varying disc
outer radius for different values of the disc aspect ratio power law,
$s$. Note that the timescale does not depend on the scaling of the disc aspect ratio, $(H/R)_{0}$, only the power law, $s$.  The larger the disc outer radius, the shorter the
timescale. This is because closer to the neutron star perturber, the KL oscillation
timescale is shorter 
(see equation~(\ref{time})). The timescale is relatively insensitive to the value  of $s$, but the smaller its value, the more
mass that is located farther out in the disc and thus the shorter the
oscillation timescale. 

The observed timescale between two subsequent type II outbursts is around
$3\,\rm yr$.  The  KL oscillation timescales predicted by the initially circular orbit disc model are
 longer than the observed  outburst timescales except for very large truncation radius and for a constant disc aspect ratio ($s=0$).  We note that the error in the masses of the binary components may lead to a decrease of a few tens of percent in the timescale (see Section~\ref{binary}). This is not sufficient to explain this discrepancy. However,  we showed in Section~\ref{test} that the KL oscillation timescale may be up to a factor of a few shorter for an initially eccentric disc. Thus, in order to reproduce the timescale between two subsequent type II outbursts, there must be eccentric orbit material left in the disc after the outburst. This is in agreement with our finding in Section~\ref{tvisc} that the disc cannot be completely destroyed during the outburst as the viscous timescale in the disc is long.  

The isothermal ($s=0.5$) 3D hydrodynamical simulation described in \cite{Martinetal2014} had a peak
eccentricity at a time of about $22\,\rm P_{\rm orb} \approx 1.5 \,\rm yr$. Since this is half of the KL oscillation timescale, $\left<\tau_{\rm KL}\right>/2$,  this appears to be consistent with the recurring type II outburst timescale observed, despite the fact that the disc was initially circular. However, the disc was able to undergo the first KL oscillation on an artificially short timescale due to the initial surface density profile which had $\Sigma \propto R^{-1}$. Since there is no source of mass in the disc simulation, the surface density evolves  towards a steady state accretion disc with $\nu \Sigma$=const which gives $\Sigma \propto R^{-3/2}$ for an isothermal accretion disc.  The isothermal decretion discs described in this work have $\Sigma \propto R^{-2}$. Thus, the initial disc set up had artificially too much mass in the outer parts of the disc and this led to a shorter KL oscillation timescale.   A second outburst was not seen in the simulation because there was no material added at the disc inner edge and the resolution of the disc became too poor after the first outburst. In the future we will model the evolution of a disc that includes the accretion of material at the inner edge from the star. Three dimensional simulations would take into account not only the surface density evolution, but non--axisymmetric spiral arms \citep[e.g.][]{Panoglou2018,Panoglou2019} that cannot be included into our one--dimensional approach. However, the advantage of our analytic model is that we can explore a wide range of parameters.

Simulations of accretion discs undergoing KL oscillations previously
have shown damped oscillations that decay in time. The inclination of
the disc gradually decreases until the inclination is below the
critical KL angle
\citep[e.g.][]{Martinetal2014b,Fu2015,Fu2015b}. However these
simulations did not have any source of material to the disc that would keep the inclination of the disc high. We
suggest that the Be star disc can continue to undergo KL oscillations
provided that the spin of the Be star remains misaligned to the
orbital plane and thus provides a source of highly inclined material
to the disc. Damping of the spin--orbit misalignment is expected to occur on the same timescale as rotation synchronisation \citep{Hut1981,Eggleton2001}. The synchronisation timescale \citep{Hurley2002} is much longer than the lifetime of the Be star except for the smallest orbital period Be/X-ray binaries \citep{Stoyanov2009}.  Thus we expect that a spin--orbit misalignment imparted by the supernova that formed the neutron star to last for the lifetime of the Be/X-ray binary. 
 
\section{{Disc KL oscillation} timescale dependence on binary orbital period}
\label{porb}

\begin{figure} 
\begin{centering} 
\includegraphics[width=8.5cm]{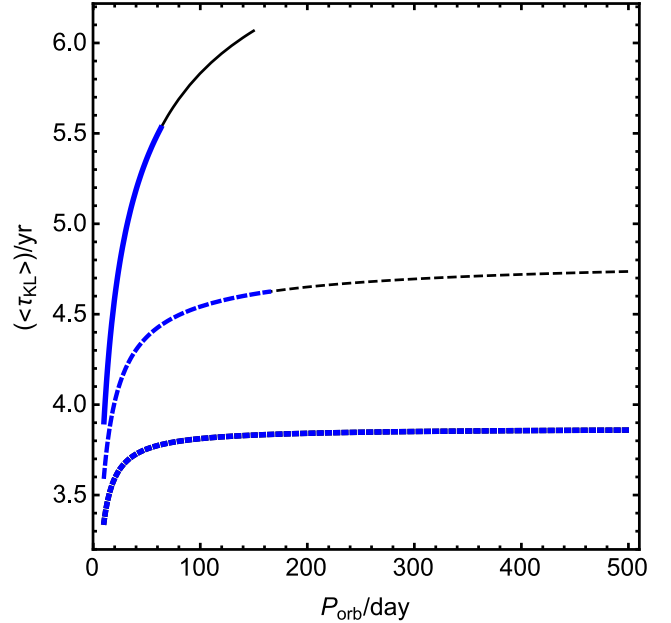} 
\end{centering} 
\caption{The KL oscillation timescale for an initially circular steady state decretion disc with
  surface density distributed as equation~(\ref{sigma}) between the
  stellar radius $R_{\rm in}=8\,\rm R_\odot$ and $R_{\rm t}=50/95a$ for
  $s=0.5$ (solid lines), $s=0.25$ (dashed lines) and $s=0$ (dotted lines). The thick blue lines have $(H/R)_0=0.08$ and the thin black lines have $(H/R)_0=0.06$. The lines are truncated where $H/R=(H/R)_{\rm crit}$. The blue lines lie exactly on top of the black lines.}
\label{fig:period} 
\end{figure}

The models presented so far have focused on the orbital parameters of the Be/X-ray binary system 4U 0115+634. We now examine how the mechanism operates for parameters relevant for Be/X-ray binary systems with different orbital period. 
Fig.~\ref{fig:period} shows the KL oscillation timescale for a disc in a system with varying orbital period. We assume that the disc is truncated at $R_{\rm t}=(50/95)\,a$, which is relevant for binary eccentricity $e_{\rm b}=0.34$ and disc inclination of $i=70^\circ$. We vary the semi--major axis $a$ such that the orbital period changes in the range $10{\,\rm day}<P_{\rm orb}<500\,\rm day$. Note that smaller binary eccentricities  and larger disc inclinations lead to wider discs \citep{Artymowicz1994,Lubowetal2015,Miranda2015,Brown2019} and hence shorter KL oscillation timescales. Generally, the KL oscillation timescale is quite insensitive to the orbital period of the binary, within a factor of 2, assuming that the disc has expanded out to the tidal truncation radius.  The viscous timescale at the outer edge of the disc increases with binary separation. This viscous timescale exceeds the KL oscillation timescale for longer orbital period binaries.

The flared disc model lines in Fig.~\ref{fig:period} are truncated where $H/R (R=R_{\rm t})=(H/R)_{\rm crit}$. 
For the isothermal disc model, we estimate the critical binary orbital period below which the Be star disc is unstable to the KL mechanism. We require that the disc aspect ratio (Eq. \ref{eq:aspectratio} with $s=0.5$) is less than or equal to $(H/R)_{\rm crit}$ at the tidal truncation radius and find
\begin{equation}
   P\lesssim P_{\rm crit}\approx 150\, \left(\frac{\eta}{50/95}\right)^{-3/2}\,\left(\frac{(H/R)_0}{0.06}\right)^{-3}
    \,\rm day,
    \label{eq:pcrit}
\end{equation}
where $(H/R)_0$ is the disc aspect ratio at $R_0=50\,\rm R_\odot$ and we defined the ratio $\eta=R_{\rm t}/a$. For longer binary orbital period, $P>P_{\rm crit}$, (i.e. higher aspect ratios) the disc is stable against KL oscillations.   We have assumed that the disc is wide and viscous enough to reach its tidal truncation radius. If the disc does not expand out to the truncation radius, type~II outbursts are not likely to occur.
Therefore equation~(\ref{eq:pcrit}) represents a necessary but not sufficient condition for the disc to be KL unstable and thus for type II outbursts to occur.

Note that this critical period should also depend on both the eccentricity and the inclination of the binary orbit but we have only taken that into account through the parameter $\eta$. If the binary is less eccentric, or the disc is more inclined, the size of the disc relative to the binary separation increases, i.e. $\eta$ increases \citep{Artymowicz1994,Lubowetal2015}. This leads to a smaller critical orbital period since the more extended disc has larger H/R in the disc outer parts and is more stable to KL oscillations.  

In summary, if the disc has a flared vertical structure,  the disc aspect ratio at the outer edge may exceed the critical value required for it to be unstable to KL oscillations in systems with long binary orbital periods. Thus, KL oscillations are more likely in smaller orbital period binaries. 
This is in agreement with observations that show that type~II outbursts are more likely to occur in short period binaries \citep{Cheng2014}.  The critical orbital period depends on the disc aspect ratio and the disc radial extent, which depends on the eccentricity of the binary orbit, the inclination of the disc and the mass ratio of the binary.

\section{Discussion}
\label{discussion}

If KL disc oscillations drive type~II outbursts, then any Be star X-ray binary is expected to have type~II
outbursts providing that two conditions are satisfied. First, the
Be star decretion disc must be misaligned to the binary orbital
plane by more than the critical angle required for KL
oscillations. For a test particle in an initially circular orbit around one component of a circular orbit binary, this is $39^\circ$. However this angle varies for thick decretion discs depending upon the properties of the disc \citep{Lubow2017}. 
Secondly, the orbital period of the binary must be short enough to allow the Be star disc to expand sufficiently far to overflow its Roche lobe during a KL oscillation while still being unstable to KL oscillations (see equation~(\ref{hcrit})). This is in agreement with the observational data provided in \cite{Cheng2014} that show that Be/X-ray binaries with short orbital period  are more likely to undergo type II outbursts. However, there are some observed systems with long orbital periods that exhibit type II outbursts. GRO~J1008-57 has an orbital period of  $250\,\rm days$ and shows giant outbursts \citep[e.g.][]{Kuhnel2017}. For this system to  undergo KL oscillations, the disc aspect ratio must be close to constant with radius for the mechanism to operate. 
Swift J1626.6-5156, with an orbital period of $132\,\rm days$ \citep[e.g.][]{Reig2011b}, and 1A 0535+262, with an orbital period of $111\,\rm days$ \citep{Acciari2011} also show giant outbursts. In order for the KL disc mechanism to drive giant outbursts in these systems the disc must not be very flared. The disc aspect ratio must scale $H/R\propto R^{s}$ with $s\lesssim 0.25$. The KL disc driven giant outburst mechanism could be ruled out through observations of a disc with a low misalignment or a high disc aspect ratio.

A large portion of the Be star disc may be destroyed
during the KL oscillation. The material can be re-accreted on to the Be-star (because the disc is highly eccentric),
transferred to the neutron star or to a circumbinary orbit  \citep{Franchini2019}, or be
ejected from the binary star system. However, the disc
continues to be fed 
at the inner edge from the Be star. The eccentricity of the disc evolves due to the KL mechanism and due to the addition of circular orbit material at the inner edge.
The KL oscillations occur on a shorter timescale with smaller amplitude for a larger initial particle eccentricity.  Thus in the disc, this leads to a shorter timescale
before the next outburst. Since this leaves less time for material to
accumulate in the outer parts of the disc, we expect a smaller outburst, as is
observed for closely spaced outbursts  in 4U~0115+634 \citep[e.g.][]{Reig2018}. Thus,
in a disc the KL oscillations depend sensitively on how much material
is left over after the outburst, the eccentricity of the material and the decretion rate of material into the disc from the star.

The amount of material that is transferred from the Be star disc to the neutron star depends on several parameters. In particular \cite{Franchini2019} found that for a globally isothermal accretion disc ($s=0.5$) the amount of material around the secondary star was smaller compared to the case with a constant disc aspect ratio ($s=0$). Also, a higher disc aspect ratio as well as higher initial circumprimary disc inclination leads to larger amount of material being transferred. 
Furthermore the lower the binary mass ratio, the higher the mass transfer to the secondary star. This is relevant to this work because of the extreme mass ratios of these systems. We expect the mass transfer to be very high in Be star X-ray binaries undergoing KL oscillations and with Be star discs filling the Roche lobe.   

The inclination of the disc also has a significant affect on the amount of mass transfer. The higher the initial disc inclination, the stronger the eccentricity growth and the more mass that is transferred to the companion. For an inclination of $70^\circ$ (but for an equal mass binary), most of the material in the Be star disc was transferred to the companion \citep[see Figure 10 in][]{Franchini2019}.

\section{Conclusions} 
\label{concs}

We have found that KL oscillations of the Be star disc in the Be/X-ray binary 4U~0115+364 operate on a timescale that matches the observed frequency of type II (giant) X-ray outbursts provided that the disc is not completely destroyed during the outburst. A highly inclined disc
around a Be star becomes eccentric and overflows its Roche lobe so
that material flows on to the companion neutron star. 
The KL oscillation timescale for a disc depends sensitively
on the distribution of material within the disc and the eccentricity of the material.  Since the disc is
depleted during a giant outburst, material must be replenished between
outbursts. The viscous timescale must be short enough for the disc to spread outwards between outbursts. For 4U~0115+634 this requires the disc aspect ratio at the outer disc edge to be $H/R \gtrsim 0.06$. However, the disc aspect ratio at the outer disc edge must be small enough for the KL mechanism to operate, $H/R \lesssim 0.11$. KL mechanism driven type~II X-ray outbursts are more likely to occur in shorter period binaries, if the disc is flared.

\section*{Acknowledgements} 
 
We thank an anonymous referee for a thorough review. We acknowledge support from NASA through grant NNX17AB96G.


\label{lastpage} 
\end{document}